\documentclass[11pt]{article}
\usepackage{moriond,epsfig}

\def\be{\begin{equation}}
  \def\ee{\end{equation}}
\def\bea{\begin{eqnarray}}
  \def\eea{\end{eqnarray}}

\newcommand{\afig}{{\kern -0.25em a}}
\newcommand{\bfig}{{\kern -0.25em b}}
\newcommand{\cfig}{{\kern -0.25em c}}
\newcommand{\dfig}{{\kern -0.25em d}}

\newcommand{\MW}{M_{\mathrm{W}}}
\newcommand{\MT}{M_{\mathrm{t}}}
\newcommand{\MH}{M_{\mathrm{H}}}
\newcommand{\GW}{\Gamma_{\mathrm{W}}}

\newcommand{\OA}{\cal{O}(\alpha)}
\newcommand{\TO}{\rightarrow}
\newcommand{\EE}{\mathrm{e^+e^-}}
\newcommand{\WW}{\mathrm{W^+W^-}}

\newcommand{\FFFF}{\mathrm{ffff}}

\newcommand{\QQP}{\mathrm{q} \overline{\mathrm{q^\prime}}}
\newcommand{\QQ}{\mathrm{q} \overline{\mathrm{q}}}
\newcommand{\LN}{\ell \nu}

\newcommand{\QQLN}{{\mathrm{q q} \ell \nu}}
\newcommand{\QQQQ}{\mathrm{q q q q}}
\newcommand{\LNLN}{\ell \nu \ell \nu}

\def\GeV{\ifmmode {\mathrm{\ Ge\kern -0.1em V}}\else
  \textrm{Ge\kern -0.1em V}\fi}%
\def\MeV{\ifmmode {\mathrm{\ Me\kern -0.1em V}}\else
  \textrm{Me\kern -0.1em V}\fi}%
\def\keV{\ifmmode {\mathrm{\ ke\kern -0.1em V}}\else
  \textrm{ke\kern -0.1em V}\fi}%
\def\eV{\ifmmode  {\mathrm{\ e\kern -0.1em V}}\else
  \textrm{e\kern -0.1em V}\fi}%
\begin{document}
\vspace*{4cm}
\title{MEASUREMENT OF THE W BOSON MASS AT LEP}

\author{ ARNO STRAESSNER }

\address{DPNC, University of Geneva, \\
  24 Quai Ernest-Ansermet, 1211 Geneva 4, Switzerland \\
  e-mail: Arno.Straessner@cern.ch}

\maketitle\abstracts{ The mass of the W boson is measured at LEP by
  fully reconstructing the W boson decays. The measurement techniques
  and systematic uncertainties are presented. The current measurement
  of the mass of the W boson at LEP yields $80.412\pm 0.042 \GeV$.}

\section{Introduction}
At LEP, W bosons are produced in the reaction $\EE \TO \WW$ with the
subsequent decay of the W's into quark pairs, $\QQP$, or a lepton and
a neutrino, $\LN$. About 40000 W pair events are registered by 
the four experiments ALEPH, DELPHI, L3 and OPAL, corresponding to a
total luminosity of $2.8$ $\mathrm{fb}^{-1}$.

One of the main goals of the LEP programme is to determine the mass of
the W boson, $\MW$, from the reconstructed invariant mass spectra.
Involved techniques are used to obtain an optimal statistical
precision. However, in the fully hadronic channel, systematic
uncertainties are important, like final state interactions (FSI) between the
hadronically decaying W bosons. These uncertainties reduce the
sensitivity of this channel. The recent activities concentrate on
increasing the weight of the hadronic events to obtain a globally more
precise result on the LEP W mass.

\section{Extraction of the W Mass}

\begin{figure}
  \begin{center}
    \parbox{0.49\linewidth}{
      \begin{center}
        \includegraphics[width=0.3\linewidth]{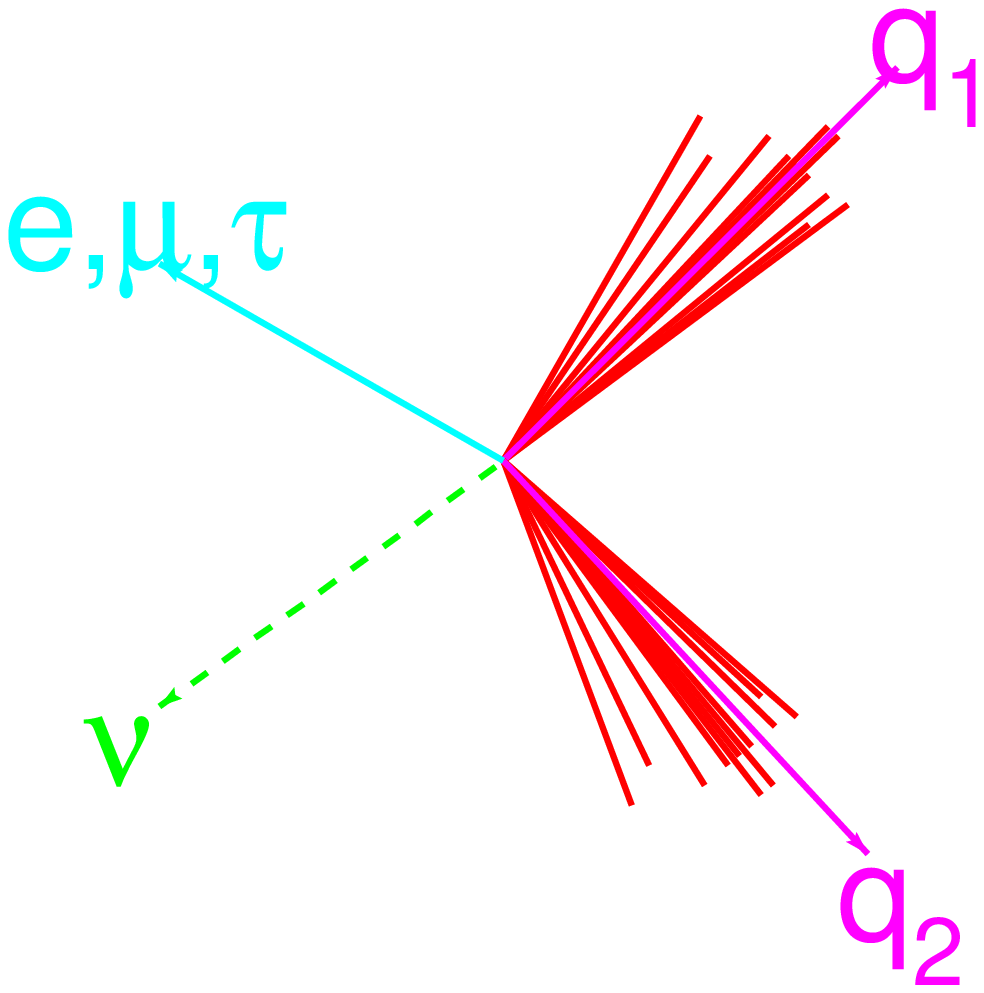} 
        \hspace*{0.5cm}
        \includegraphics[width=0.3\linewidth]{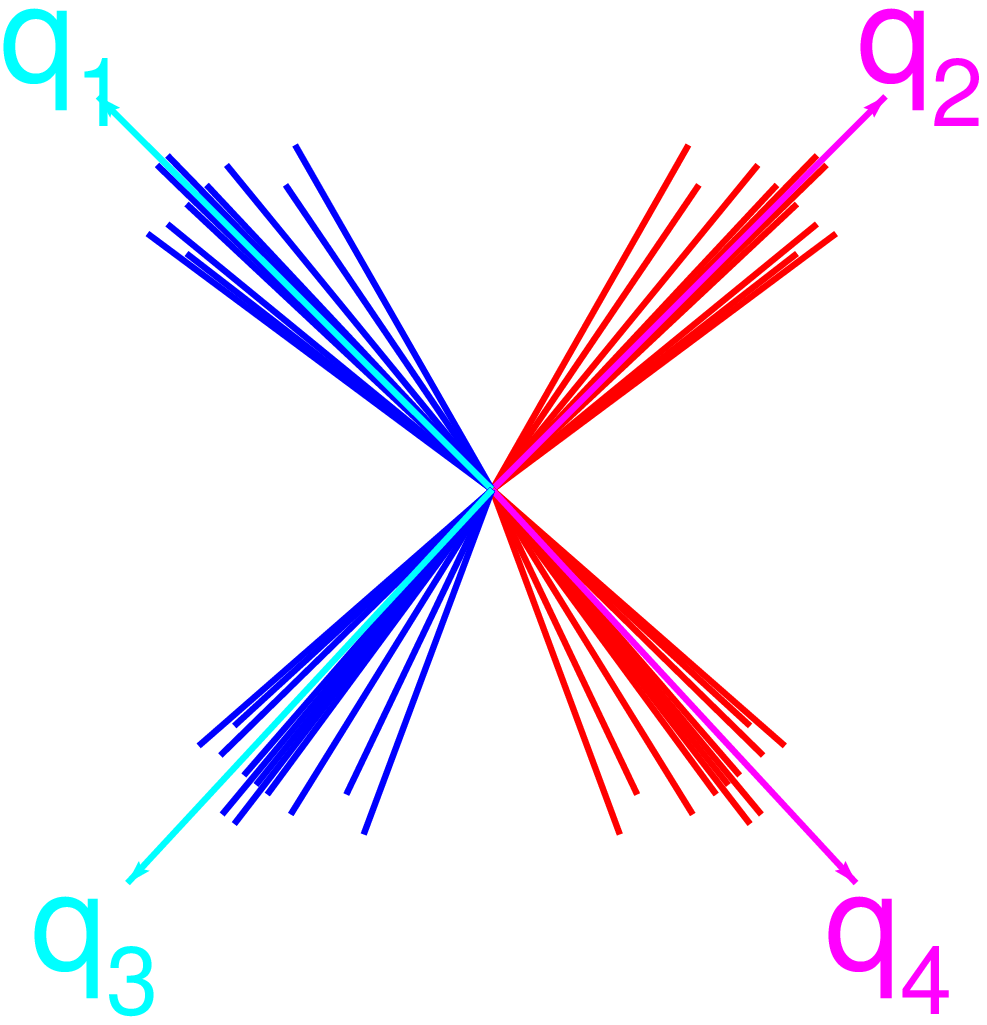} \\
        \includegraphics[width=0.3\linewidth]{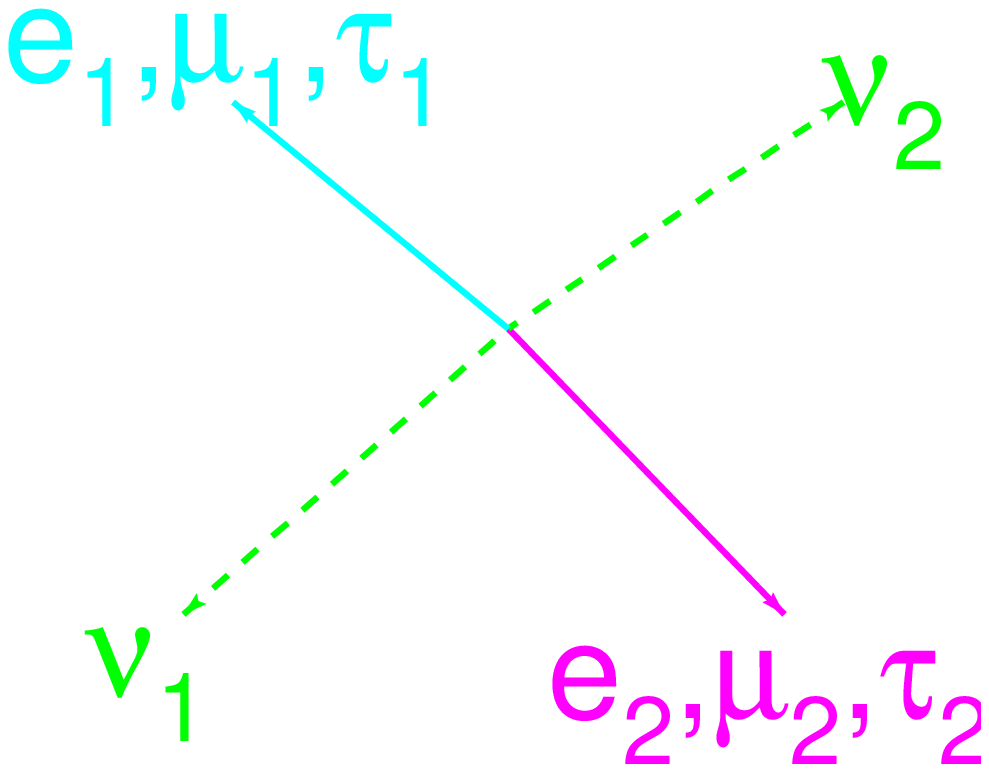} 
      \end{center}
      }
    \parbox{0.49\linewidth}{
      \includegraphics[width=\linewidth]{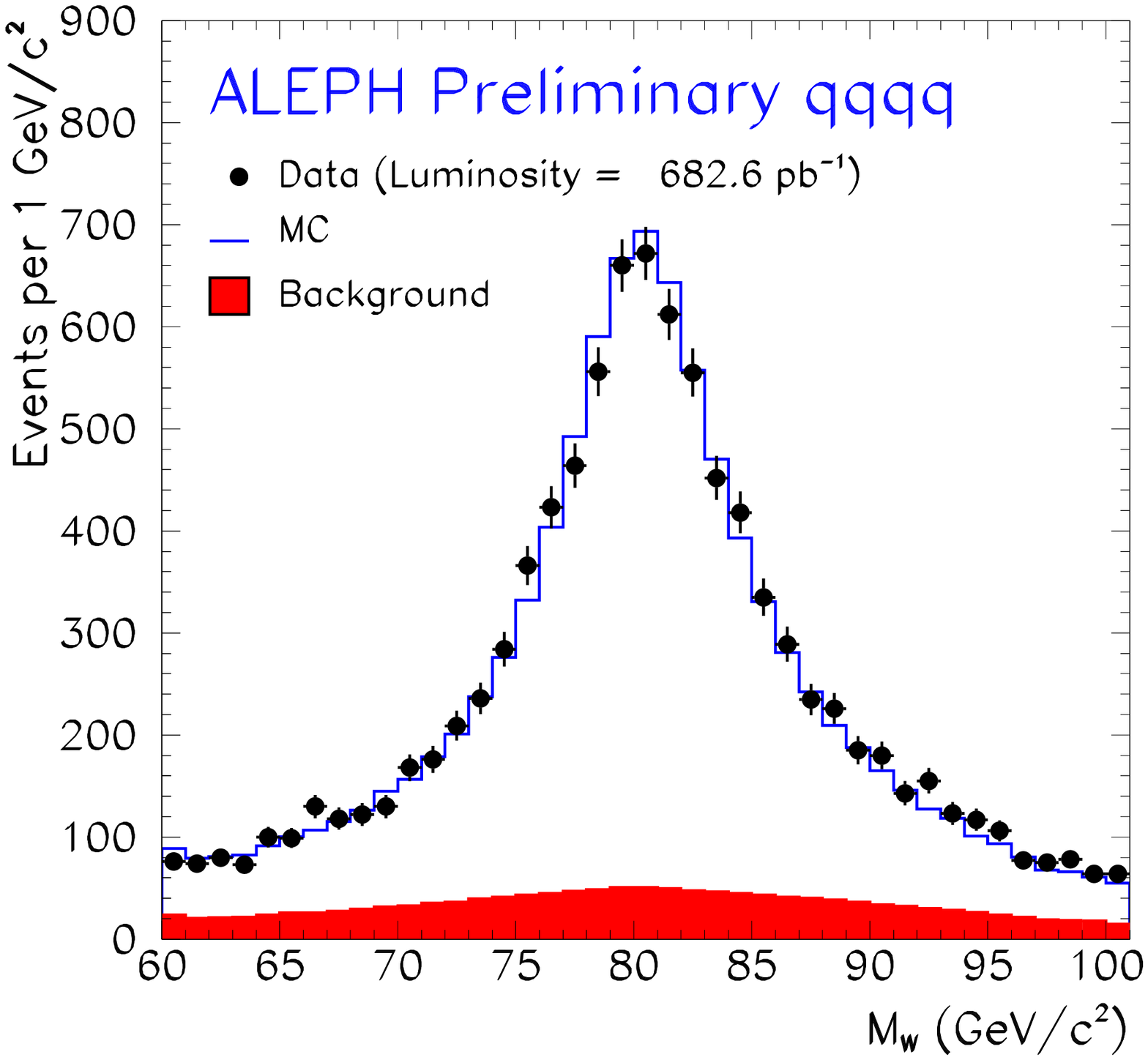}
      }
    \caption{a) The three four-fermion final states in W pair
      production, $\QQQQ$, $\QQLN$, $\LNLN$, are reconstructed. b)
      Invariant mass spectrum in the $\QQQQ$ final state as measured
      with the ALEPH detector.
      \label{fig:spectrum}}
  \end{center}
\end{figure}

In the semi-leptonic and fully hadronic WW decay channels the complete
four-fermion final states, shown in Figure~\ref{fig:spectrum}~\afig,
can be reconstructed. The most sensitive observable for the W mass
measurement is the invariant mass of the decaying W bosons. To improve
the resolution on this quantity, a kinematic fitting procedure is
applied. The reconstructed four-momenta of the final state fermions
are varied within their resolution and kinematic constraints, like
energy-momentum conservation and equal invariant masses in each event,
are imposed. In fully hadronic events the resolution is diluted due to
the different pairing combinations of the final state jets. In
general, the most probable pairings according to the kinematics are
chosen.  Figure~\ref{fig:spectrum}~\bfig\ shows an example of the
invariant mass spectrum measured in the $\QQQQ$ channel.

In $\LNLN$ final states the event kinematic can not be reconstructed
completely and the leptonic energy spectrum and a pseudo-mass are
chosen as optimal W mass estimators.

To obtain W mass and width, three basic extraction methods are used:
in the Monte Carlo reweighting procedure, the underlying $\MW$ and
$\GW$ values of the Monte Carlo prediction are varied and compared to
the measured spectra; in the convolution method the theoretically
predicted spectra are folded with resolution functions and fitted to
data; in the Breit-Wigner method the resonance curve is fitted to data
and possible bias is calibrated with Monte Carlo samples.  All methods
exploit information from several event observables, including the
invariant masses obtained in kinematic fits with different constraints,
as well as their uncertainties.

With the current techniques an equal statistical precision of $32\MeV$
and $35\MeV$ is reached in the $\QQLN$ and $\QQQQ$ channel,
respectively.

\section{Systematic Uncertainties}

A common source of systematic uncertainty in the $\QQLN$ and $\QQQQ$
channels is the theoretical description of photon radiation that is
used in the Monte Carlo simulation. In total, the uncertainty on
initial state radiation (ISR), final state radiation (FSR) and $\OA$
electroweak corrections amounts to $8\MeV$ in all channels. The
current numbers are based on reweighting of Monte Carlo events
according to possible differences in the theoretical prediction of the
photon spectrum.  However, recent comparisons between two different
Monte Carlo generators, YFSWW~\cite{yfsww} and RacoonWW~\cite{racoon},
show larger differences. The uncertainty is therefore expected to be
underestimated and may increase to $10-15\MeV$ for the final LEP W
mass measurement.

Another common systematic error source is the description of the
hadronisation of quarks. The uncertainty is mainly derived from the
difference between the hadronisation models Pythia, Herwig and
Ariadne~\cite{frag-models}. Especially the simulated rate of heavy
baryons influences the jet masses and therefore the derived invariant
masses. A reweighting to the measured baryon rates is performed by
Opal. Delphi also compares mixed Lorentz boosted Z decays, that are
arranged to reproduce W-pair decay kinematics, to the hadronisation
models. Since the agreement between the different models and data have
improved, the current systematic uncertainty of $18-19\MeV$ may be
reduced in future.

\begin{figure}
  \begin{center}
    \includegraphics[width=0.49\linewidth]{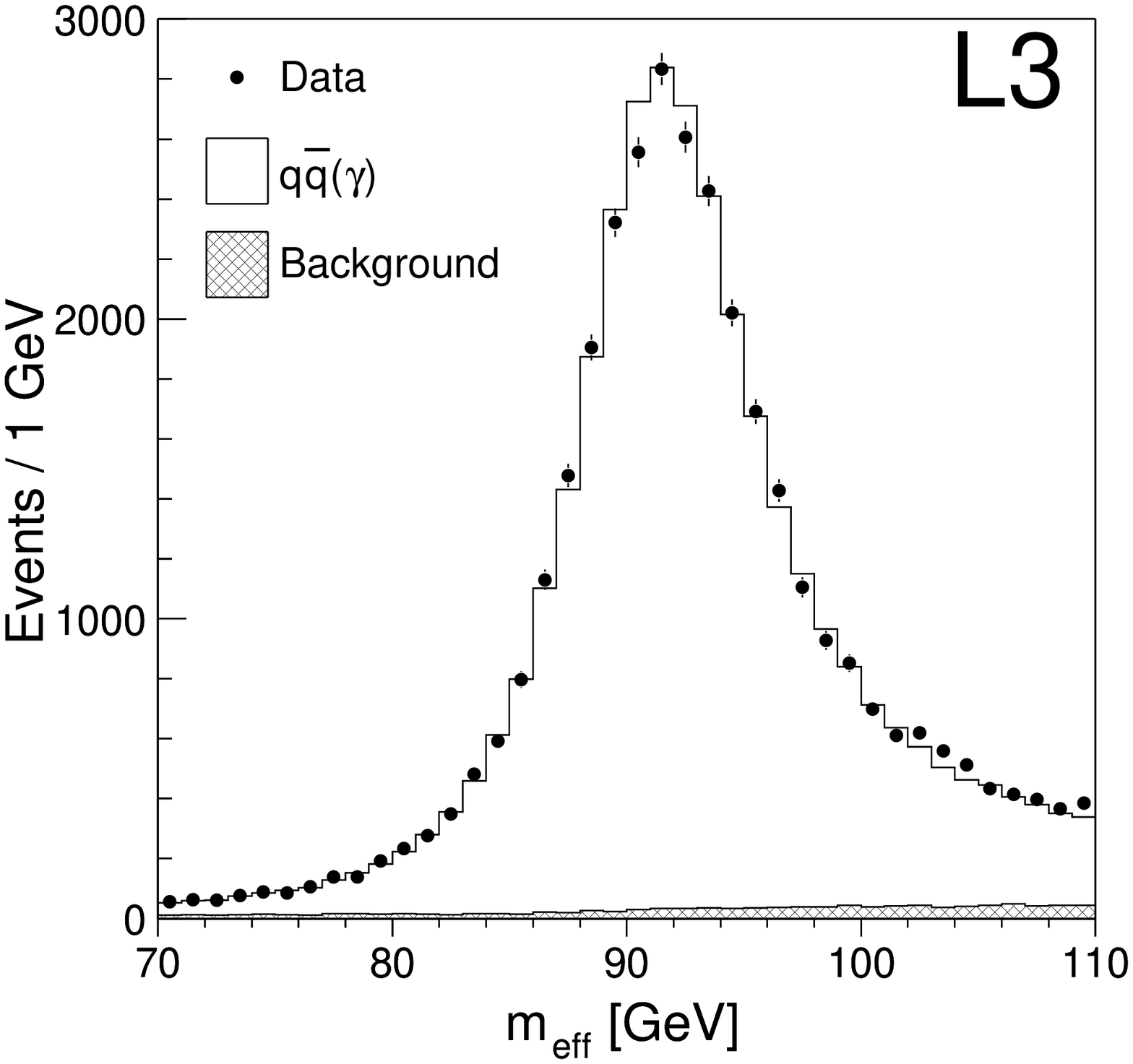} 
    \includegraphics[width=0.49\linewidth]{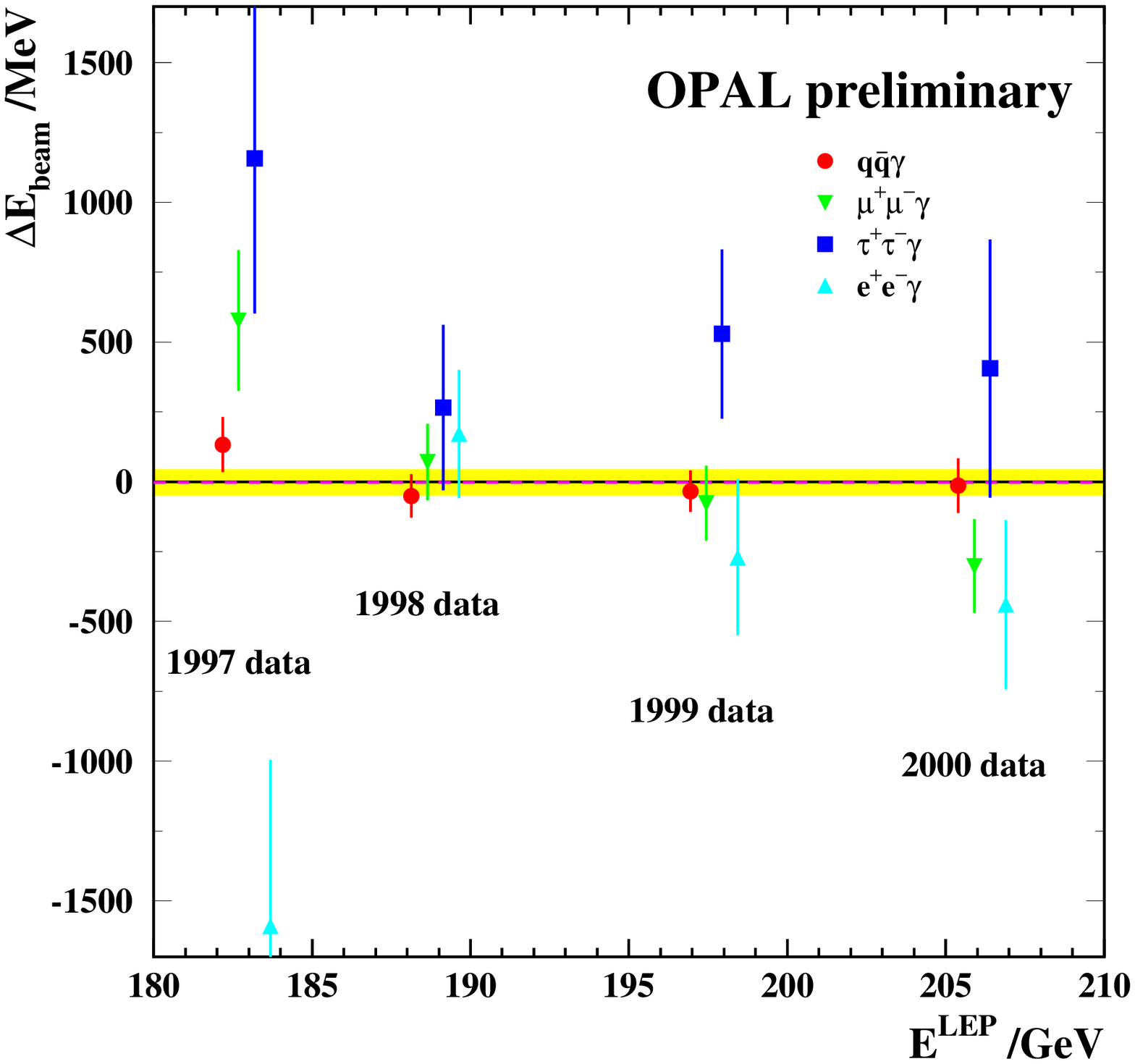}
    \caption{a) Invariant Z mass spectrum reconstructed in $\QQ\gamma$
      events. b) Difference between the LEP beam energy and the beam
      energies determined by Opal in an analysis of radiative
      fermion-pair events. \label{fig:zret}}
  \end{center}
\end{figure}

A similar reduction of systematic uncertainty, which is not yet
included in the LEP combined W mass value, is originating from the LEP
beam energy uncertainty. The final LEP energy calibration has
improved~\cite{elep} with respect to the current calibration applied.
It will result in a much smaller uncertainty of only $10\MeV$, with
only a small change of the average LEP beam energies. The LEP
experiments also performed a cross-check of the LEP energy measurement
by reconstructing the Z boson mass in radiative fermion-pair events,
as shown in Figure~\ref{fig:zret}. Both Opal and L3 obtain results in
good agreement with the LEP energy calibration~\cite{zret}.

Z decays at high centre-of-mass energies as well as in Z peak
calibration data are analysed to test the detector simulation. Energy
and angular measurement of leptons and jets in data are compared to
Monte Carlo simulation. Possible differences are corrected and the
remaining uncertainty on the difference is taken as systematic
uncertainty, which amounts to $14\MeV$ on the combined W mass result.

\begin{table}[t]
  \begin{center}
    \begin{tabular}{l | c | c | c }
      Source & \multicolumn{3}{c}{Uncertainties  on $\MW$ in $\MeV$} \\
      & \ \ $\QQLN$ \ \ & \ \ $\QQQQ$ \ \ & \ \ $\FFFF$\ \ \ \\
      \hline
      ISR/FSR               &      8 &       8 &          8 \\
      Hadronisation       &   19 &    18 &       18 \\
      LEP Beam Energy       &     17 &      17 &         17 \\
      Detector Systematics        &     14 &      10 &         14 \\
      Colour Reconnection &   -- &    90 &        9 \\
      BE Correlations     &   -- &    35 &        3 \\
      Other                 &      4 &       5 &          4 \\
      \hline
      Total Systematic             &   31 &   101 &       31 \\
      \hline
      Statistical                  &   32 &    35 &       29 \\
      \hline
      Total                   &    44 &     107 & 
      43  \\
      \hline
      Statistical in absence of Systematics &   32 &    28 &       21 \\
    \end{tabular}
    \caption{Systematic uncertainties in the W mass measurement.\label{tab:syst}}
  \end{center}
\end{table}

\section{Final State Interactions}

A special complication in reconstructing the invariant masses appears
in the fully hadronic channel. Hadronic FSI may introduce cross-talk
between the two decaying W bosons of one event. Bose-Einstein
Correlations (BEC) lead to an increased production of identical
bosons, like pions and kaons, close in phase space. Colour
Reconnection (CR) changes the colour flow between the four quarks in
the final state. By reconnecting colour strings between the previously
colour-neutral di-quark systems from each W decay, momentum is
transferred between the W's and the hadronisation process is altered.

As listed in Table~\ref{tab:syst}, BEC and CR may change the
reconstructed W mass by $35\MeV$ and $90\MeV$, respectively. The
increase of CR effects with centre-of-mass energy is taken into
account.

At LEP, FSI effects are measured also in other observables, which are
used to constrain the various BEC and CR models. The BEC measurements
in W pairs mainly concentrate on charged pion production.  Assuming a
spherical and gaussian shaped source emitting the pions, the
two-particle correlation function $C$ can be written as:
\begin{eqnarray*}
  C(Q_{\pi\pi}) = 1 - \lambda \exp(-R^2 Q^2_{\pi\pi}) \; ,
\end{eqnarray*}
where $Q_{\pi\pi}$ is the square of the four-momentum difference of
the two pions, $-(p_{\pi,1}-p_{\pi,2})^2$. The parameter $\lambda$ is
the correlation strength and $R$ is the inverse radius of the source.
When analysing semi-hadronic W events, it is found that the BEC
between pions coming from the same W boson agree very well with the
correlations observed in Z decays, if $\mathrm{Z}\TO\mathrm{bb}$ is
suppressed. 

\begin{figure}
  \begin{center}
    \includegraphics[width=0.45\linewidth]{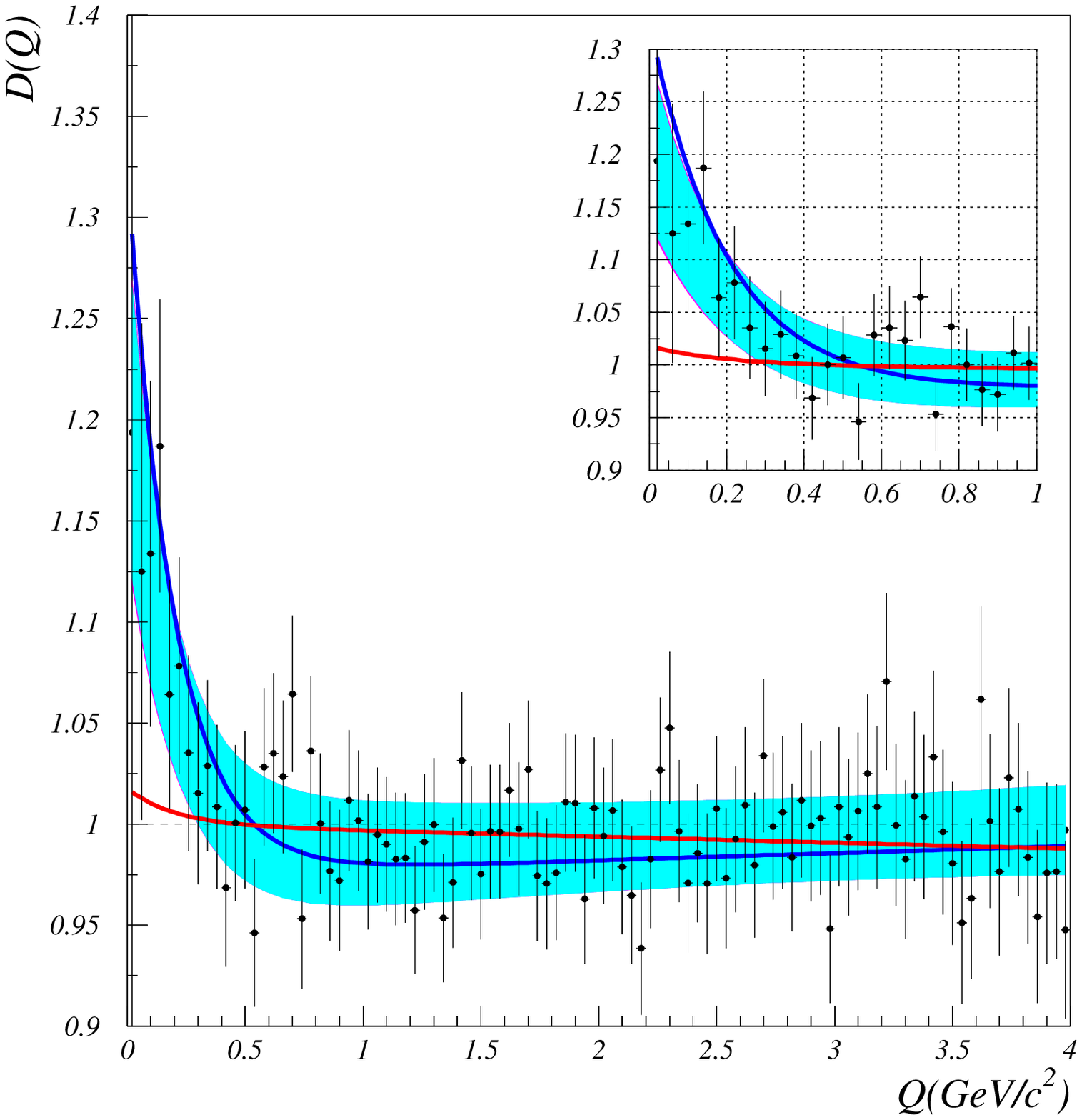} 
    \hspace*{0.5cm}
    \includegraphics[width=0.35\linewidth]{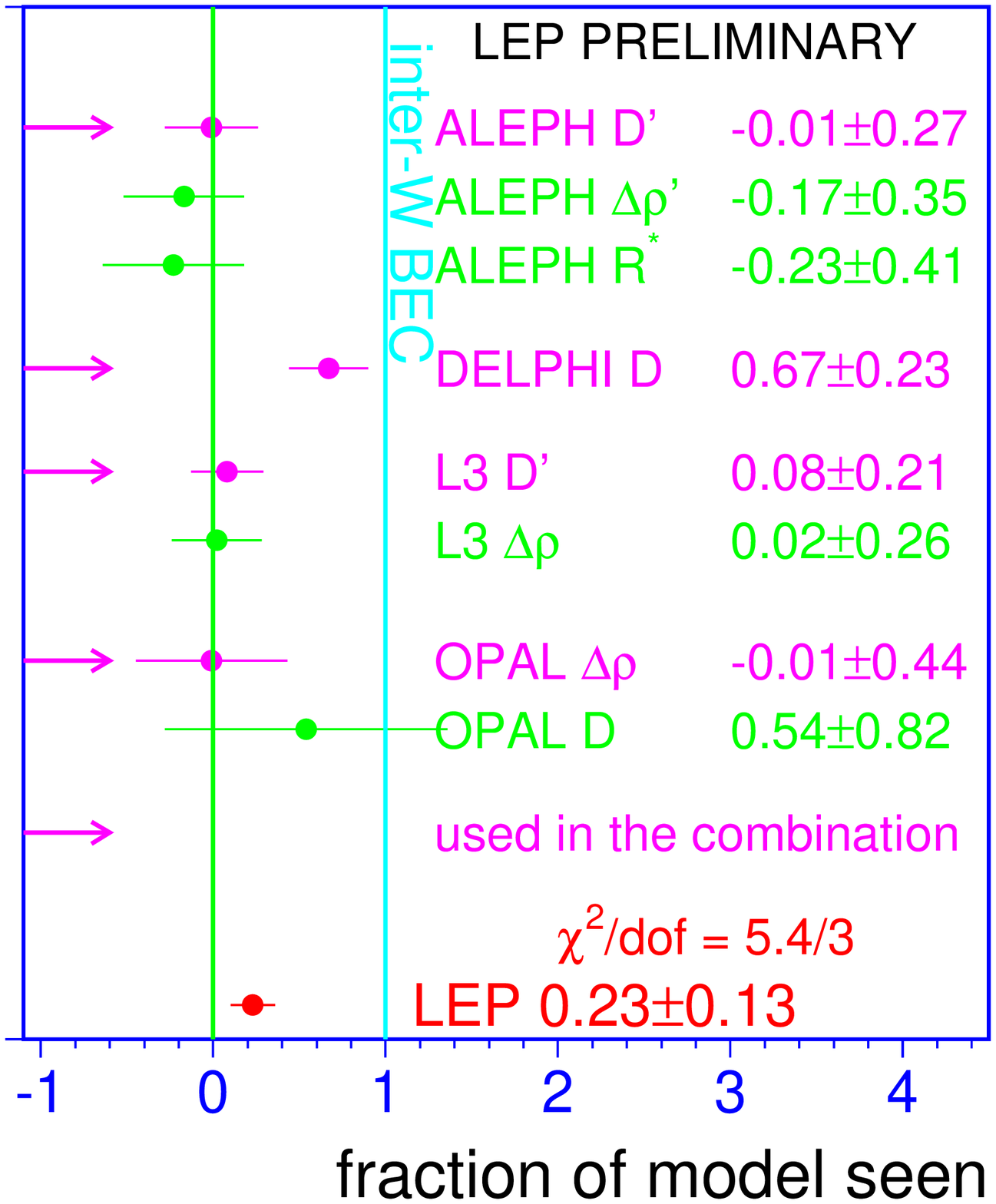}
    \caption{a) The Delphi measurement of the $D$ parameter shows a
    preference for BEC, visible as an enhancement of $D$ for low
    $Q$. b) BEC results from the LEP experiments and their
    combination. The individual results are normalised such that the
    maximal BEC effect yields a value of 1 and the absence of BEC a
    value of 0. \label{fig:bec}}
  \end{center}
\end{figure}

Important for the mass measurement are, however, the correlations
between pions from different W bosons. If there are no such
correlations the two-particle density in fully hadronic events can
be split into three terms:
\begin{eqnarray*}
  \rho^{\mathrm{WW}} (1,2) = \rho^{\mathrm{W+}}(1,2) + \rho^{\mathrm{W-}}(1,2) +
2 \rho^{\mathrm{W^+}}(1)\rho^{\mathrm{W^-}}(2) \; .
\end{eqnarray*}
The first two terms on the right-hand side of the equation are the
density functions for pions coming from the same W. They can be
determined in semi-hadronic events. The last term describes the case
when one pion comes from one W boson and the second from the other W
boson. This part can be constructed from a sample of mixed
semi-hadronic event: $\rho^{\mathrm{W^+}}(1)\rho^{\mathrm{W^-}}(2) =
\rho^{\mathrm{W^+W^-}}_{\mathrm{mix}}$. If the equation holds, i.e. in
absence of BE correlations between two W's, the following ratio and
difference of densities
\begin{eqnarray*}  
  D &=& \frac{\rho^{\mathrm{WW}}}{2 \rho^{\mathrm{W}} + 
    2 \rho^{\mathrm{W^+W^-}}_{\mathrm{mix}}} \\
  \Delta\rho &=& \rho^{\mathrm{WW}}- 2 \rho^{\mathrm{W}} -2 
  \rho^{\mathrm{W^+W^-}}_{\mathrm{mix}}
\end{eqnarray*}
are equal to 1 and 0, respectively.  Figure~\ref{fig:bec}~\afig\ shows
the quantity $D$ as a function of $Q$ as measured by Delphi. The
Delphi data are consistent with moderate BEC between pions from
different W's. However, the combination of all LEP experiments prefers
the absence of those correlations~\cite{ewwg-note}, as shown in
Figure~\ref{fig:bec}~\bfig.  Monte Carlo studies show that the
corresponding W mass shift is proportional to the BEC strength, so
that the systematic uncertainty on $\MW$ due to BEC can in future be
reduced to $10-15\MeV$ using the direct BEC measurement.

\begin{figure}
  \begin{center}
    \includegraphics[width=0.49\linewidth]{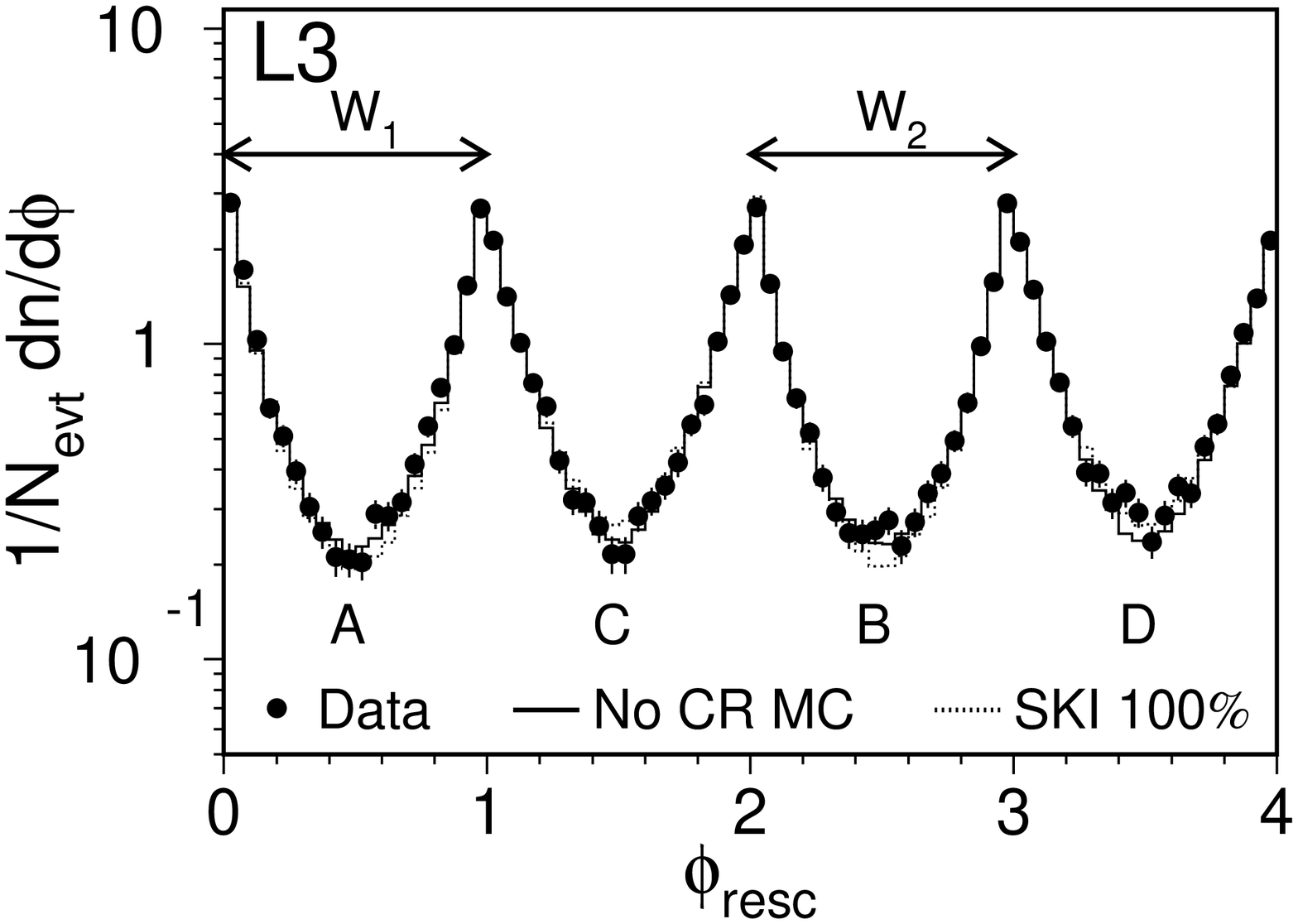}
    \includegraphics[width=0.49\linewidth]{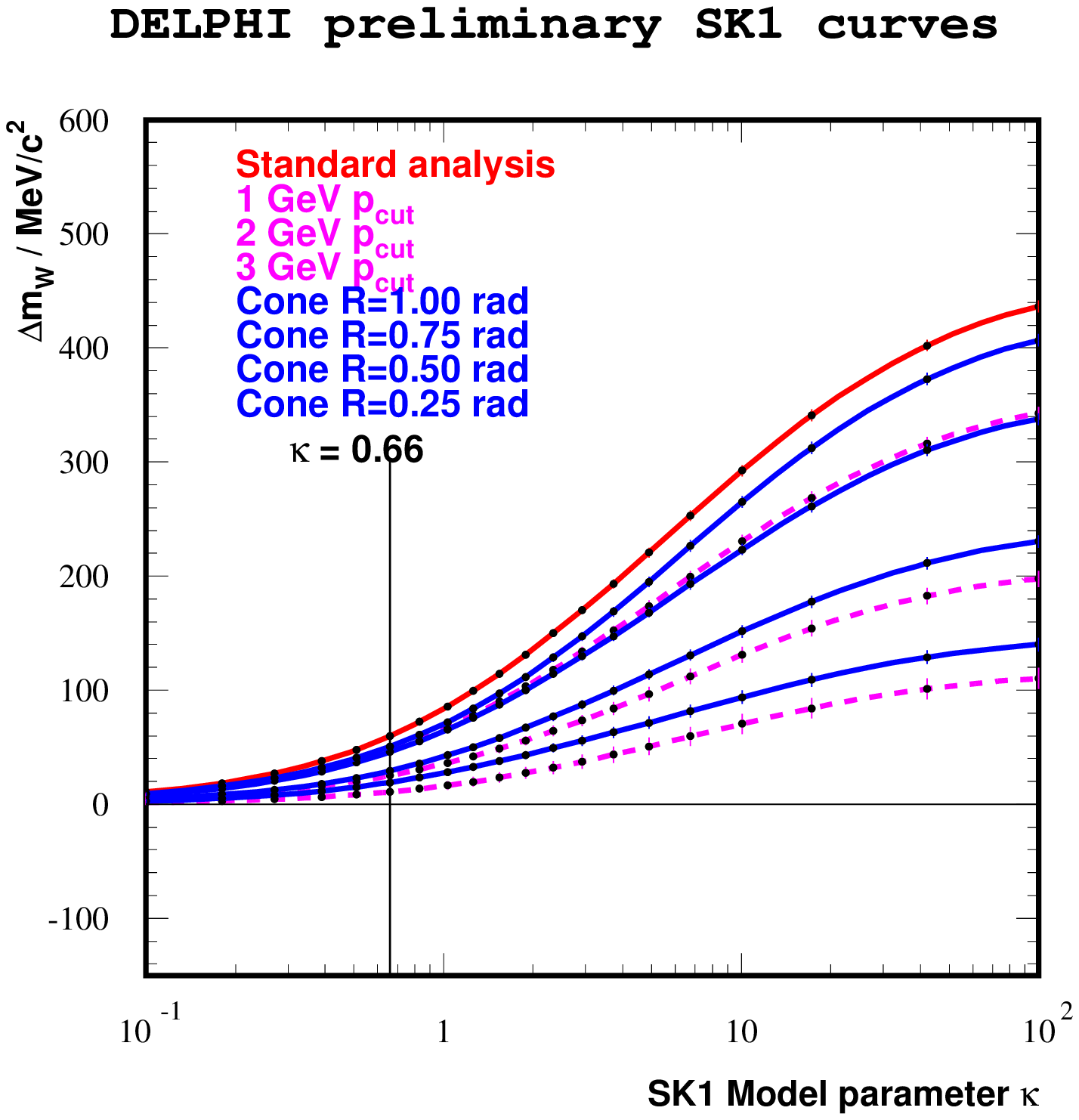} 
    \caption{a) Particle flow measured by L3 in the regions inside the
      W's (A.B) and between W's (C,D). b) W mass bias due to CR for
      different jet reconstruction methods. Restricting the jet cone
      size and the energy threshold reduces the CR effect.
      \label{fig:cr}}
  \end{center}
\end{figure}

A similar approach is made in the reduction of the CR uncertainty.
Different models~\cite{cr} predict CR, for example the
Sj{\"o}strand-Khose (SK) models 1 and 2, the Herwig model, or the
Ariadne models 1-3. CR mainly manifests in a distortion of the W mass
distribution.  Effects on particle multiplicities are too small to be
detected in data~\cite{aleph-ch}.  The only other observable sensitive
to CR is the particle flow in the regions between the quark jets.
Figure~\ref{fig:cr}~\afig\ shows the angular distribution of particles
in $\QQQQ$ events, after rescaling the angle of the particle to the
next jet in such a way that each of the four jets is positioned at an
integer angular value.  A ratio $R$ is then calculated between the
particle flow of the regions that connect two jets of one W (A+B) and
the region that connects two jets of different W's (C+D):
\begin{eqnarray*}
  R=\frac{d\phi/dN(A+B)}{d\phi/dN(C+D)}
\end{eqnarray*}
To quantify the difference in the two regions A+B and C+D, the $R$
distribution is integrated. A variable $r$ is then constructed as the
ratio of integrals $R_N$ obtained in data and in a Monte Carlo without
CR:
\begin{eqnarray*}
  r(\mathrm{data})=R_N(\mathrm{data})/R_N(\mathrm{MC\ witout\ CR})
\end{eqnarray*}
Figure~\ref{fig:cr}~\bfig\ shows the result of the LEP experiments,
using all data. In case all LEP experiments are combined, the SK1
model with 100\% reconnection would yield a value of
$r(\mathrm{SK1-100\%})=0.891$, while a value of
$r(\mathrm{data})=0.969\pm0.015$ is obtained in the LEP measurement.
Data therefore excludes strong CR effects as predicted by SK1 with
5.2$\sigma$. The preferred values of the SK1 model parameter $k_i$ are
in the range $[0.39,2.13]$~\cite{cr-lep}.
The upper boundary of $k_i$ is used to estimate the systematic
uncertainty on $\MW$ due to CR. Averaged over all centre-of-mass
energies one yields a possible mass shift of $90\MeV$. This shift is
larger than those observed in other models, like Ariadne 2 or Herwig.

A further constraint on the CR effects can be put by the mass
measurement itself. An indication that CR can not be strong comes from
the comparison of the W mass measured in $\QQQQ$ and $\QQLN$ events:
\begin{eqnarray*}
  \Delta\MW&=&\MW(\QQQQ)-\MW(\QQLN)=+22\pm43\MeV \;,
\end{eqnarray*}
where systematic errors due to possible FSI are removed. The observed
value is compatible with zero.

In addition, the W mass bias due to CR can also be reduced by
modifying the jet algorithms that are used to reconstruct the W decay
products.  When reducing the cone size of the jets or by restricting
the energy range of the objects clustered to a jet, the CR mass bias
can be reduced, as shown in Figure~\ref{fig:cr}~\bfig.  Recent
investigations have shown that this is the case for all CR models
used. If one now varies the cone size or the energy cut-off, the mass
shift observed in data and for the CR models can be compared. This
result can also be combined with the particle flow measurement since
the correlations between the measurements are small~\cite{delphi-cr}.

The four LEP experiments foresee to perform such an alternative mass
analysis with reduced sensitivity to CR effects. This will enhance the
weight of the $\QQQQ$ channel in the LEP combination, which is
currently only 0.10.

\section{Results}

\begin{figure}
  \begin{center}
    \includegraphics[width=0.35\linewidth]{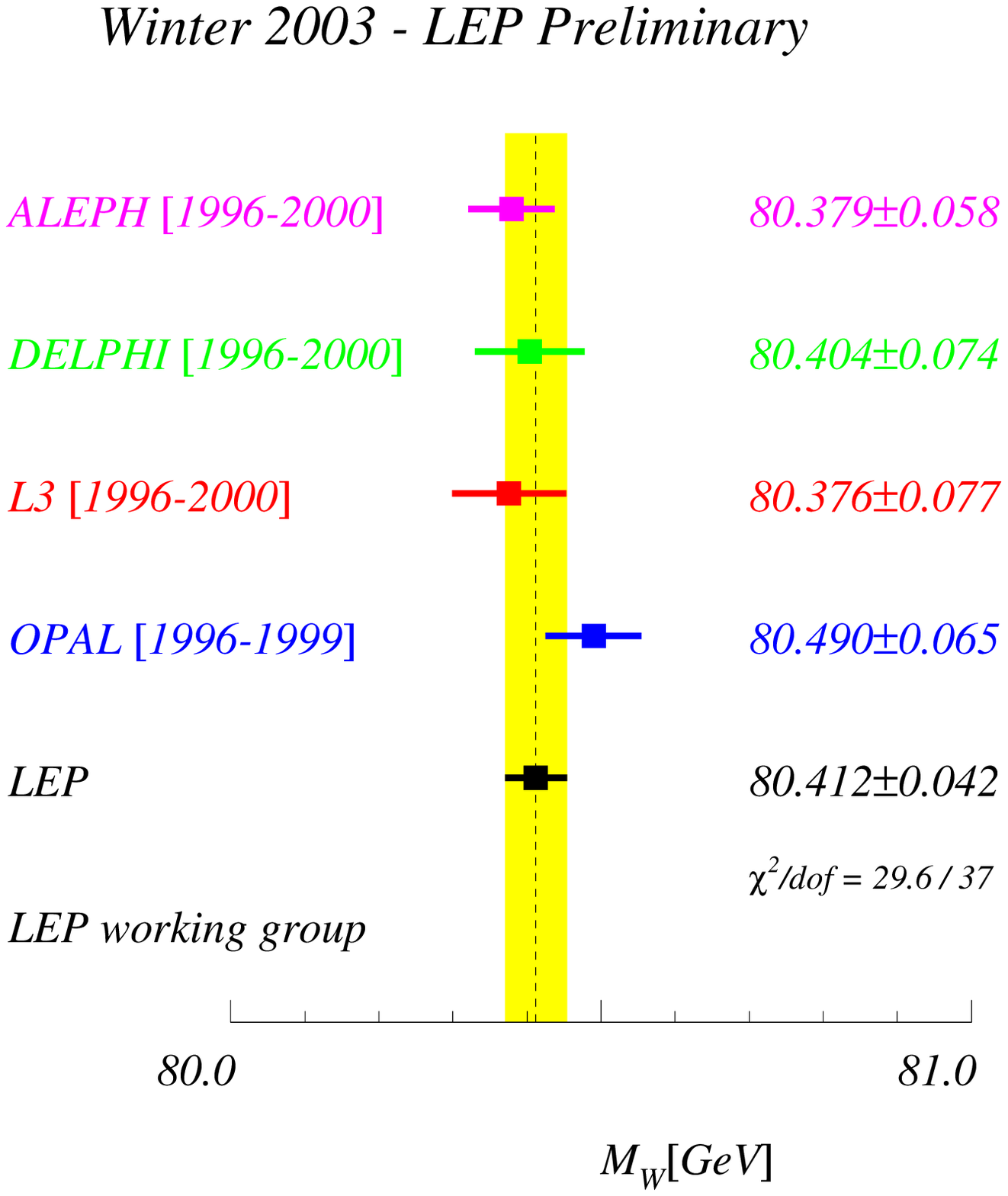} \hspace*{1cm}
    \includegraphics[width=0.40\linewidth]{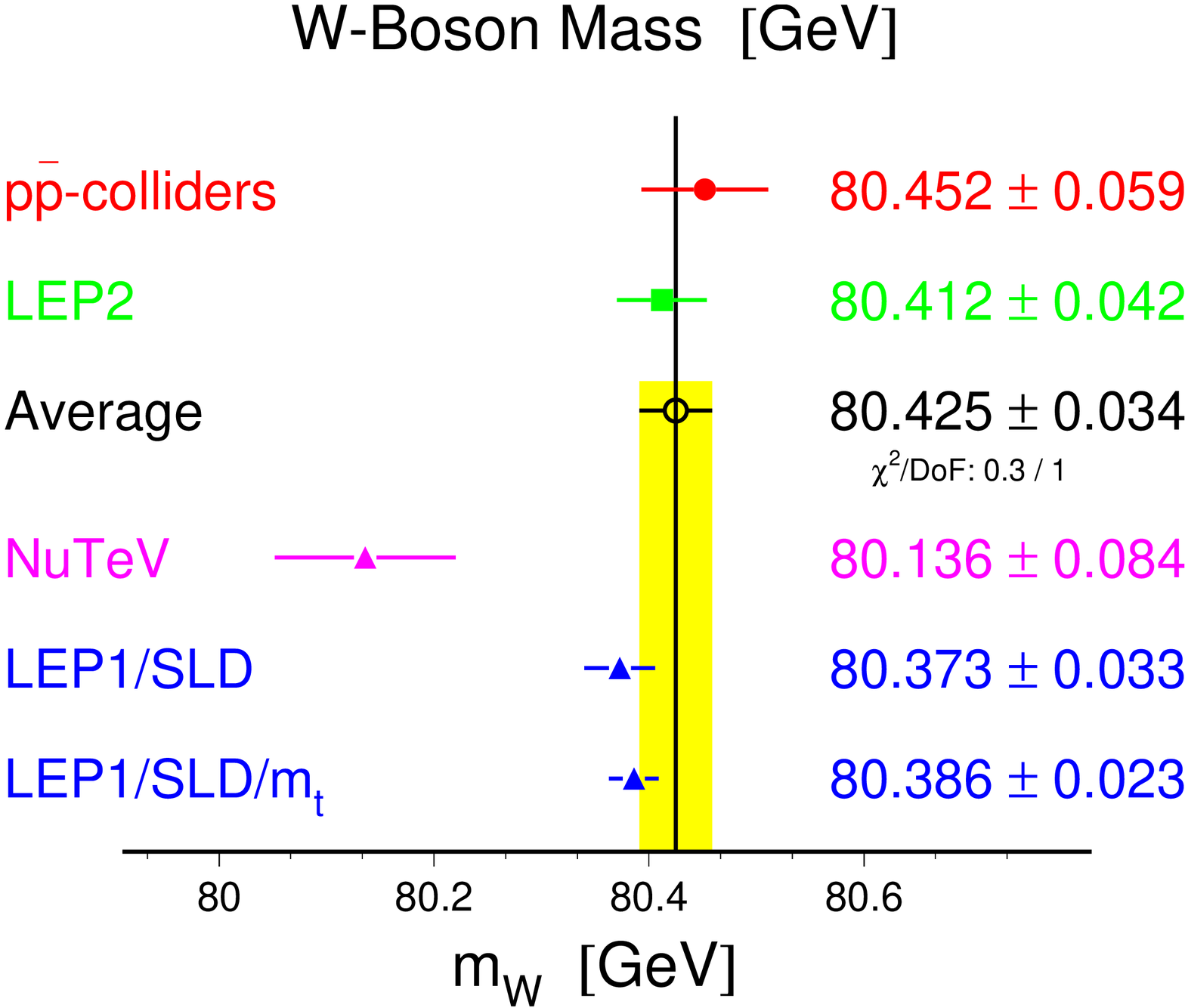}
    \caption{a) Measurements of $\MW$ by the four LEP experiments, using a
      common systematic uncertainty for FSI. b) LEP result compared
      and combined with the measurements in pp-collisions. The direct
      measurement of $\MW$ is in good agreement with the indirect
      result obtained in fits to the remaining electroweak
      data. \label{fig:res}}
  \end{center}
\end{figure}
Figure~\ref{fig:res}~\afig\ shows the individual results obtained by the LEP
experiments. With the current systematic errors and all correlations
properly included, the LEP W mass measurement split into $\QQLN$ and
$\QQQQ$ channel yields~\cite{lep-mw}:
\begin{eqnarray*}
  \MW(\QQLN)&=&80.411\pm0.032\mathrm{(stat.)}\pm0.030\mathrm{(syst.)}\GeV \\
  \MW(\QQQQ)&=&80.420\pm0.035\mathrm{(stat.)}\pm0.101\mathrm{(syst.)}\GeV \; ,
\end{eqnarray*}
with a correlation coefficient of 0.18. The combined mass value for
all channels is 
\begin{eqnarray*}
  \MW(\FFFF)&=&80.412\pm0.029\mathrm{(stat.)}\pm0.031\mathrm{(syst.)}\GeV \; ,
\end{eqnarray*}
with a good $\chi^2/\mathrm{d.o.f}$ of 28.2/33. Including the result
derived from the cross-section measurement at the W-pair production
threshold does not change numerically
the above result:
\begin{eqnarray*}
  \MW^{\mathrm{LEP}}&=&80.412\pm0.042\GeV \; .
\end{eqnarray*}

The method of direct reconstruction is also well suited to measure the
width of the W boson, $\GW$. A combined fit to the LEP data
yields~\cite{lep-mw}:
\begin{eqnarray*}
  \GW^{\mathrm{LEP}}&=&2.150\pm 0.068\mathrm{(stat.)}\pm0.060\mathrm{(syst.)}\GeV \; ,
\end{eqnarray*}
and agrees well with the Standard Model (SM) prediction of $\GW=2.099$
using the LEP W mass cited above.

As shown in Figure~\ref{fig:res}~\bfig, the LEP W mass agrees well
with the measurement at $\mathrm{p\bar{p}}$
colliders~\cite{ewwg-note}.  The combination of these direct $\MW$
measurements is also in good agreement with the indirect determination
from the other electroweak data~\cite{ewwg-note-new}. The result
obtained in $\nu$N scattering by NUTEV, which is derived from the
measurement of the electroweak mixing angle, $\sin^2\theta_w$,
deviates from the LEP result by 2.9 $\sigma$.  However, there is no
systematic effect found that may explain this difference.

\begin{figure}
  \begin{center}
    \includegraphics[width=0.49\linewidth]{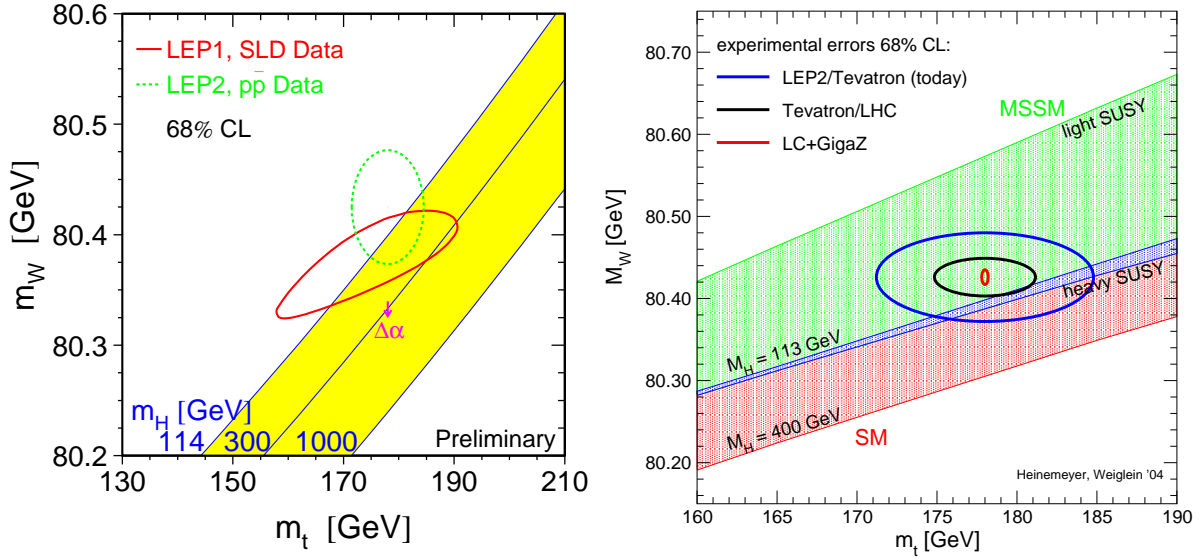} 
    \includegraphics[width=0.49\linewidth]{MWMT04.cl.eps}
  \end{center}
  \caption{ a) Contour curves for the direct and indirect measurement of
    $\MW$ and $\MT$ compared to the SM prediction for different values
    of $\MH$. b) The direct $\MW$ and $\MT$ measurements compared to
    the MSSM prediction. The exclusion limits from searches for new
    particles at LEP are taken into account in the calculation.  The
    MSSM and SM areas overlap in the region where the mass of the SM
    Higgs boson is in the MSSM range, {\it i.e.} for
    $\MH\stackrel{<}{\sim}130\GeV$.
    \label{fig:contours} }
\end{figure}

The W mass is an important parameter in the Standard Model. The
precise measurement of $\MW$ probes the SM at the level of its
radiative corrections. A comparison of the direct and indirect W mass
determinations~\cite{ewwg-note-new} is shown in
Figure~\ref{fig:contours}~\afig\ in the plane of W mass and top quark
mass, $\MT$, that is measured at the Tevatron. Good agreement between
the two sets of measurements is observed. Also shown is the SM
prediction for various values of the mass of the Higgs boson, $\MH$.
The measurements prefer small values of $\MH$.

In a supersymmetric extension of the theory, the Minimal
Supersymmetric Standard Model (MSSM), the preferred $\MW$-$\MT$ area
differs slightly from the SM prediction.  Additional radiative
correction terms, involving supersymmetric particles, shift the
$\MW$-$\MT$ band to larger values of $\MW$. The MSSM
prediction~\cite{mssm-weiglein} is shown in
Figure~\ref{fig:contours}~\bfig. The experimental precision of the
electroweak measurements is not accurate enough to decide between the
two models. However, an increased precision is expected from
measurements at the Tevatron, the future LHC and linear colliders. The
electroweak data gives also motivation to continue the search for the
Higgs boson in a low mass range, as the last missing piece of the SM,
and for supersymmetric particles at the Tevatron and at LHC
experiments.

\section{Conclusion}

The mass of the W boson is measured at LEP to
$\MW^{\mathrm{LEP}}=80.412\pm0.042\GeV$ and the width to
$\GW^{\mathrm{LEP}}=2.150\pm0.091\GeV$. It is an important
contribution to the tests of the Standard Model and its supersymmetric
extensions. Exploiting the recent results on systematic uncertainties,
the $\MW$ measurement is expected to reach a final accuracy of about
$35\MeV$, completing the many precision tests of the SM performed at
LEP.

\section*{Acknowledgements}
I would like to thank the experiments ALEPH, DELPHI, L3 and OPAL for
making their most recent and preliminary results available. I also
like to thank the LEP Electroweak and WW Working Groups as well as
Georg Weiglein and collaborators for preparing their results in form
of nice graphs and plots.

\normalsize

\end{document}